\documentclass[12pt,preprint]{aastex}

\slugcomment{Submitted to the ApJ, August 2002; Revised November 2003}

\shorttitle{Companions to Isolated Ellipticals}
\shortauthors{Madore et al.}

\begin{document}


\title{Companions to Isolated Elliptical Galaxies: \\
    Revisiting the Bothun-Sullivan (1977) Sample}


\author{Barry F. Madore\altaffilmark{1} and Wendy L. Freedman}
\affil{Observatories of the  Carnegie Institution of Washington \\ 813 Santa Barbara St., Pasadena, CA ~~91101}
\email{barry@ipac.caltech.edu, wendy@ociw.edu}

\and

\author{Gregory D. Bothun}
\affil{Department of Physics, 
415 Willamette Hall \\
University of Washington,
Eugene, OR 97403}
\email{nuts@moo.uoregon.edu}


\altaffiltext{1}{Co-Director,  NASA/IPAC Extraglactic  Database (NED),
California Institute of Technology, Pasadena, CA 92215}


\begin{abstract}

We investigate the number of physical companion galaxies for a sample
of relatively isolated elliptical galaxies.  The {\it NASA/IPAC
Extragalactic Database} (NED) has been used to reinvestigate the
incidence of satellite galaxies for a sample of 34 elliptical
galaxies, first investigated by Bothun \& Sullivan (1977) using a
visual inspection of Palomar Sky Survey prints out to a projected
search radius of 75 kpc. We have repeated their original investigation
using data cataloged data in NED.  Nine of these ellipticals appear to
be members of galaxy clusters: the remaining sample of 25 galaxies
reveals an average of +1.0 $\pm $ 0.5 apparent companions per galaxy
within a projected search radius of 75~kpc, in excess of two
equal-area comparison regions displaced by 150-300~kpc.  This is
significantly larger than the +0.12 $\pm$ 0.42 companions/galaxy found
by Bothun \& Sullivan for the identical sample.  Making use of
published radial velocities, mostly available since the completion of
the Bothun-Sullivan study, identifies the physical companions and
gives a somewhat lower estimate of +0.4 companions per elliptical.
This is still a factor of 3$\times$ larger than the original
statistical study, but given the incomplete and heterogeneous nature
of the survey redshifts in NED, it still yields a firm lower limit on
the number (and identity) of physical companions.  An expansion of the
search radius out to 300~kpc, again restricted to sampling only those
objects with known redshifts in NED, gives another lower limit of 4.5physical companions per galaxy.  (Excluding five elliptical galaxies
in the Fornax cluster this average drops to 3.5 companions per
elliptical.)  These physical companions are individually identified
and listed, and the ensemble-averaged radial density distribution of
these associated galaxies is presented.  For the ensemble, the radial
density distribution is found to have a fall-off consistent with $\rho
\propto R^{-0.5}$ out to approximately 150~kpc.  For non-Fornax
cluster companions the fall-off continues out to the 300-kpc limit of
the survey.  The velocity dispersion of these companions is found to
reach a maximum of 350~km/sec at around 120~kpc, after which
they fall at a rate consistent with Keplerian. This fall-off may then
indicate the detection of a cut-off in the mass-density distribution
in the elliptical galaxies' dark-matter halo at $\sim$100~kpc.

\end{abstract}




\section{Introduction}

Recently, the numbers of  satellite galaxy companions to galaxies have
received renewed  interest in  the context of  cold dark  matter (CDM)
models for  structure formation  where the theoretical  expectation of
the models  exceeds the observed companions by  approximately a factor
of 100  ({\it e.g.,}  Moore {\it  et al.}  1999;  Klypin {\it  et al.}
1999).   This large  number of  expected companions  results  from the
relative  inefficiency of  the  galaxy amalgamation  process from  CDM
seeds in which each large mass ($10^{12}~M_{\sun}$) is surrounded by a
large relic  population.  Given the  nature of large  scale structure,
this  relic population  can either  be confined  to  the group/cluster
potential, or  bound to individual  large galaxy halos.   In addition,
the  incidence  of  physical   companions  can  provide  an  important
constraint  on the  merger history  of galaxies  and/or the  role that
dynamical  friction plays  in adding  additional material  to galaxies
after they have formed.  The  lack of any observational counterpart to
the theoretical expectation may be  somewhat troubling but there are a
host  of viable  explanations for  their absence.   For  instance, its
possible that  the systems are just  too diffuse to  be detected, that
they  have  been  essentially  heated/evaporated  by  an  ionizing  UV
background and/or supernova winds due to star formation in the massive
galaxies.

Some years ago, Bothun \& Sullivan (1977; hereafter BS77) selected out
a sample of 34 elliptical galaxies  for a survey of their dwarf galaxy
companions.  These ellipticals were defined to be isolated, based on a
criterion  that no  other large/bright  galaxy was  found to  be  at a
projected radius  of 150 kpc  of the elliptical.  In  retrospect, that
isolation criteria was  naive, but little was known  about large scale
structure at that  time.  Using the POSS prints  and a digitizer, BS77
claimed to have searched for companions down to a limiting diameter of
0.2 arcmin, although recognizing galaxies at that limiting diameter on
POSS prints is likely much more difficult than BS77 appreciated.  From
that survey, BS77 concluded  that, statistically speaking, this sample
of ``isolated''  elliptical galaxies  had an excess  of +0.12$\pm$0.42
companions,  {\it i.e.,} indistinguishable  from zero.   This differed
significantly when compared to a  similar study by Holmberg (1969) who
found  +1.08 $\pm$  0.37 companions  around selected  spiral galaxies.
Both  studies  utilized the  POSS  prints  and statistical  background
fields that  were selected to  lie relatively near the  target galaxy.
Again, given  our knowledge of  the extent of large  scale clustering,
the  control fields  were,  in  general, placed  too  near the  target
galaxy.

In the ensuing years the study of companions has been continued by
Zaritsky and by Vader, and each of their collaborators, in a series of
papers (Zaritsky, Smith, Frenk, \& White 1993, 1997; Lorrimer, Frenk,
Smith, White \& Zaritsky 1994; Vader \& Sandage 1991, Vader \&
Chaboyer 1997 and references therein).  Zaritsky studied the
kinematics of the general population of companions orbiting spiral
galaxies; while Vader specifically investigated the dwarf elliptical
populations around central galaxies of diverse morphological types.
Conclusions of direct relevance to this paper (see Table 1 for a
concise summary) are that nearby field spirals typically have only one
or two physical companions each, and that of those the early-type
dwarf-elliptical companions are more preferentially concentrated
around their parent galaxies.  This may be indicative of a population
of bound physical companions, similar to that seen around some cD
galaxies in clusters ({\it e.g.,} Bothun \& Schombert 1990).

More recently the Sloan Digital Sky Survey (SDSS) and the 2dF Galaxy
Redshift Survey have each resulted in parallel and complementary
investigations into the frequency and distribution of satellites
around high-luminosity galaxies in those samples. Praha et al. (2003)
report the theoretically expected fall-off of velocity dispersion with
radius for a sample of 283 SDSS primaries having at least one orbiting
companion within a projected radius of 260~kpc. And similarly,
Brianerd \& Specian (2003) report on the kinematics of 1,556 satellites
orbiting 809 isolated host galaxies drawn from the 2dF Survey. And for
completeness we also note, in passing, that weak gravitational lensing
studies also bear on the question of companions and the distribution
of dark matter around isolated galaxies, as exemplified by the study
of McKay et al. (2001).

And finally, Garcia-Barreto, Carrillo \& Vera-Villamizar (2003) chose
78 SB galaxies from the {\it Revised Shapley-Ames Catalogue} (Sandage
\& Tammann 199*) and searched for neighboring galaxies out to 20
central-galaxy diameters (corresponding to radial separations up to 1
Mpc). 347 radial-velocity-confirmed neighbors are identified and
listed, leading to an average of 4.5 neighbors per barred
galaxy. Omitting galaxies in the Virgo cluster, such as NGC~4435, 4535
and 4654, drops the average to about 3 neighbors per SB galaxy.

In this paper our interest is refocused on the companion population
surrounding elliptical galaxies, specifically the `isolated elliptical
galaxies' studied by BS77.  Although the procedures used in BS77 were
documented, none of the surrounding galaxies counted in either
investigation were individually identified or subsequently published,
since the stated intent was to draw a statistical conclusion about the
general incidence of companions, not to necessarily identify or follow
up any particular galaxy grouping or any given (apparent) companion.
The conclusion was that the environments of ellipticals and spirals
were significantly different, in the sense that isolated ellipticals
were statistically devoid of companions, while spirals had on average
one physical companion each.

In the intervening quarter century many new surveys (2MASS, APM, etc.)
to fainter limiting magnitudes have been undertaken; follow-up radial
velocity surveys (UZC, 2dF, etc.)  have been completed; and these
published data are almost all now accessible in electronically
searchable form through the {\it NASA/IPAC Extragalactic Database}
(NED).  As part of a larger, on-going program to study the general
distribution of companion galaxies, we have undertaken here to first
revisit the BS77 sample. However, given the fainter magnitude limits
and the radial velocities now available, it seemed worthwhile to
extend their investigation and to search for true physical companions,
rather than simply statistical excesses.  These additional surveys
should provide improved sensitivity to companion detection primarily
because compact galaxies, often difficult to distinguish from stars on
the POSS, are often emission line galaxies ({\it e.g.,} Salzer,
McAlpine \& Boroson 1989; Bothun {\it et al.}  1989) that are easily
detected at other wavelengths (IR, UV/X-ray).  Hence, NED offers us
the ability to perform the kind of multi-wavelength search for
companions that can greatly extend that which can be done by visual
inspection alone.

Although  NED taps many  different surveys  for galaxy  detection, the
nature of  the radial velocity  measurements for NED  galaxies remains
sparse and heterogeneous,  so in that sense NED cannot  be used to any
reliable completeness level  for companion searches.  However, despite
the patchwork nature of NED, it does have the unique characteristic of
being the  largest on-line collection of galaxies  available, with the
most  complete sampling of  radial-velocity data  (291,000 out  of 4.7
million  objects  in  NED  have published  radial  velocities)  linked
directly to the individual objects.  In addition, the accessibility of
these on-line  data allows us  to extend the BS77  companions position
search  to a  projected radius  of 300~kpc  around each  target galaxy
thereby  increasing  the  area  surveyed  by  a  factor  of  16.   All
radial-velocity companions are identified and recorded, and the source
density as a function  of radius calculated for all velocity-confirmed
companions (Table 3).  The obvious  advantage here is that the process
is no longer statistical, as in the previous sense of differencing two
large  numbers in  order  to  extract the  companion  signal from  the
background noise, but rather it is physical, and allows us to set very
strict {\it  lower} limits  on the numbers,  and the  projected radial
distances,  of individually identifiable  galaxies that  are certainly
physically associated with the `isolated' galaxies in this sample.

\section{Reworking the Bothun \& Sullivan (1977) Sample}

We  start  by repeating  the  BS77 experiment  as  closely  as can  be
reconstructed from their published procedures.  Their sample consisted
of  34 elliptical galaxies  originally drawn  from the  {\it Reference
Catalogue  of  Bright Galaxies}  (de  Vaucouleurs  and de  Vaucouleurs
1964),  falling  north  of   $\delta  =  -45$,  and  having  expansion
velocities  V $\le$  3,000~km/sec. This  yielded a  master list  of 90
ellipticals, which was narrowed down  to a sample of 34 after applying
the  aforementioned  isolation  criterion.    Again,  as  in  BS77,  a
projected  search radius  of  75~kpc was  initially adopted  (assuming
$H_o$ =  75 km/sec/Mpc and  using the expansion  velocities, corrected
for Solar motion, as originally  tabulated in BS77).  For this sample,
the  search radii  ranged  from 9  to  14~arcmin.  We  do  not have  a
diameter limit  as in  the BS77 study,  since these are  not available
within  NED;  however, our  aim  is  simply  to search  for  potential
companions,  regardless  of  diameter.   To  assess  the  field-galaxy
contamination  rate  around  each   target  galaxy,  we  searched  two
equivalent circular areas  20-40 arcmin to the north  and to the south
of  the primary position  (BS77 used  similar comparison  regions, but
chose to  place them to the  east and west of  the central elliptical.
Our north-south positioning was done for computational simplicity, but
should be statistically equivalent.)   These regions were chosen to be
close  enough  to  be  representative  of  the  local  background  but
sufficiently  far afield  so as  to minimize  the possibility  of them
still  containing  physical  companions.   At these  separations  even
objects having similar radial velocities would not have had sufficient
time to  make one orbital  crossing and are  unlikely to be  in virial
equilibrium  with   the  target  object.    Searches  were  undertaken
interactively.  The version of NED  searched had a public release date
of 19 June 2003. Table 2  lists the results.  Columns 1 and 10 identify
the elliptical  galaxy under consideration.  Column 2  gives the total
counts (in the 75~kpc circular area = N75), while columns 3 and 4 give
the field counts  ([N] and [S], to the  north and south, respectively.
The statistical excess and its associated counting error, is listed in
column  5.  For comparison  the numbers  quoted on  a galaxy-by-galaxy
basis  by BS77  are given  in column  6, followed  by the  redshift in
column   7.    Columns   8   and   9   then   give   the   number   of
radial-velocity-confirmed  companions  inside  75 and  150~kpc  search
radii, respectively (see Section III.)

We find that within the 75~kpc search radius the isolated
elliptical-galaxy sample has on average +0.74 $\pm$0.95 cataloged
companions per elliptical (All errors reported and/or plotted in this
paper are one-sigma.)  This is the result of differencing two rather
large numbers (529 objects on target minus an average of 504 galaxies
in the equal-area comparison fields), as confirmed by the large
statistical uncertainty quoted.  However, the result differs for
galaxies inside and outside of clusters. We note that several of the
galaxies (NGC~1351, NGC~1395, NGC~1426, NGC~1427 and NGC~1439) are
probably members of the Fornax cluster.  In addition, NGC~3640,
NGC~3818, NGC~4697 and NGC~6958 are each projected upon regions of the
sky where galaxy counts are particularly rich and deep, and may also
be in loose clusters themselves (the NGC~3640 Group and Abell S0900,
in particular).  Examining the remaining 25 ellipticals the source
density of companions rises in both absolute terms and in relative
significance to +0.96$\pm$ 0.54 companions per elliptical.  This is
significantly higher in density than the result quoted by BS77 of 0.12
$\pm$ 0.42 companions/galaxy.  To first order, this indicates that the
procedure used here, which accesses multiple galaxy catalogs, has a
much higher detection efficiency than the visual approach adopted by
BS77.




\section{Looking Beyond a 75~Kpc Radius}

We went through the BS77 sample a second time, increasing the search
area a factor of 16$\times$) ({\it i.e.,} out to 300~kpc, which is
equivalent to approximately half the distance from the Milky Way to
M31.)  However, we included only those galaxies with cataloged radial
velocities commensurate with them being physically associated with the
central elliptical galaxy ({\it i.e.,} generally within
$\pm$800~km/sec) were counted and retained for further analysis and
tabulation. 

Table 3 lists the parent galaxy, each of its radial-velocity confirmed
(physical)  companions,  their  radial  separations from  the  central
elliptical (in  arcmin), and their differential  radial velocities (in
the sense of companion minus central elliptical).  Physical companions
falling within  the original BS77  search radius of 75~kpc  are marked
with  an asterisk  and are  illustrated in  Figure 1--8.   The central
velocity  dispersion  of the  parent  elliptical  galaxy  is given  in
parentheses following the galaxy name.

For the entire sample of 34 E galaxies we find 154 companion galaxies
located within a projected radius of 300~kpc and having published
redshifts within $\pm$800~km/sec of the central parent galaxy. The
average number of companions is 4.5 per elliptical, but the variation
in the number of companions per parent galaxy is large (ranging from
none so far detected around NGC~0821, NGC~2974, NGC~2986, NGC 3962,
NGC~4494 or NGC~5061, to an anomalous high of 28 associated companion
galaxies in the case of NGC~1427, in the Fornax cluster). Previously
quoted statistical (counting) uncertainties do not apply here: this is
a lower limit on the number of companions in each case, with the
uncertainty dominated by cosmic variance and survey systematics.

We now examine the radial density distribution of these physically
associated galaxies.  Figure 10 shows the ensemble radial density
fall-off as a function of normalized distance ($R_{300} = R/300$~kpc).
If it were not for the fact that these galaxies are all
radial-velocity confirmed to be physically associated with the central
elliptical, it would be tempting to suggest that there is some
background (contamination) level at which the counts are going flat at
large radii.  However, these galaxies are in all probability actual
physical companions.  Therefore, it is likely that the general
population of galaxies associated with these central ellipticals could
extend even further out beyond 300~Kpc.  Alternatively put, the
elliptical galaxies may not be all that isolated to begin with and
what we may be seeing here is the background level of their
association with a more widely distributed cluster population.

To some degree this observation may explain why the BS77 number is so
low.  It is reasonable to assume that they were differencing out
physical companions by having their comparison field in so close to
the parent galaxy that they had were not yet sampling the pure field
contamination population. Given that most of the comparison fields in
BS77 had only a average of 2 galaxies counted (and a maximum of 10 in
two fields combined) the vagaries of small number statistics become
problematic.

That said, it may also be true that our adopted velocity difference is
too generous and tends to exaggerate the number of physical companions
by  being  overly  inclusive,  especially  in  the  the  outskirts  of
clusters. Indeed, the effect of  narrowing the velocity window down to
$\pm300$~km/sec results in the loss  of 51 companions from the 300~kpc
sample,  and correspondingly  drops the  average number  of `physical'
companions down to 2.8 companions per elliptical.



\section{Discussion }


Because  of the  inherent inhomogeneity  of the  NED holdings  of both
objects over the sky and radial velocities of selected subsets, little
can be  said about the  absolute total numbers of  galaxies physically
associated  with  the  `isolated'  ellipticals  in  the  BS77  sample;
however, strict lower  limits are obtained here.  That  is to say, the
galaxies found to  be radial-velocity companions now will  not go away
with deeper surveys, with  better statistics, or with more homogeneous
studies.  We  therefore conclude that  this sample of galaxies  has 14
currently  known  physically  associated  companions, amounting  to  a
lower-limit   average   of    0.4    companions/galaxy   within
75~kpc. Within a larger 300~kpc radius (16$\times$ ~larger area) these
numbers increase  to 147 galaxies around 34  central objects, yielding
at   least  4.3  companions/galaxy.    Excluding  the   five  galaxies
(NGC~1351, 1395, 1426, 1427 and  1439) in the Fornax Cluster, gives an
average of 3.5 companions per central galaxy.

It is of interest to use the satellites as probes of the halo mass and
calculate  the mass  M interior  to  the last  radial bin,  $R_{max}$.
Using  the simplest  assumptions  and assuming  a relaxed  equilibrium
state one estimate of the mass can be obtained from ~
$M = {{3R_{max}\sigma _r^2}/{2G}}$,
~where $\sigma _r$ is the observed radial-velocity dispersion and G is
the gravitational constant.  Adopting a value of 300~km/sec as
representative of the observed velocity dispersion, at a conservative
distance of 100~kpc, the derived total mass $M_{100kpc}$ is calculated
to be in excess of $3 \times 10^{12} M_{\sun}$.  This can only be
considered an indicative mass, given the wide range of possible mass
models that can be used in inverting the projected velocity
dispersion.  (In this regard the interested reader is referred to the
more extensive modeling and discussion by Zaritsky \& White
1994). However, if the fall-off in velocity dispersion with radius, as
indicated in Figure 12, is real then this may be evidence that we have
detected a truncation in the density profile of the dark-matter halo.

The radial fall-off in satellite surface density $\Sigma$ (Figure 10)
is consistent with the general correlation function (binned) down to
30~kpc, with $\Sigma \sim r^{\alpha}$ where $\alpha = -0.5$. This is
entirely consistent with the almost identical conclusion drawn by Lake
\& Tremaine (1980) for the Holmberg (1969) spiral-galaxy companion
data, from which they found that $\alpha = -0.5$ held from 40~kpc down
to scales as small as 5~kpc.  However, a significantly steeper slope
to the radial fall-off profile is quoted by Vader \& Sandage (1991),
where $\alpha = -1.22 \pm 0.05$ for r = 16--270~kpc (corrected to
$H_o$ = 75 km/sec/kpc) for early-type dwarf companions around selected
E/S0 galaxies, gleaned from a visual inspection of photographic plates
used to construct the {\it Revised Shapley-Ames Catalog} (RSA: Sandage
\& Tammann 1981) and the {\it Carnegie Atlas} (Sandage \& Bedke
(1994). 

The  analysis of BS77  predicts that  within the  entire sample  of 34
isolated ellipticals only 4 (0.12  $\times$ 34) galaxies will prove to
be physical companions. Based  on incomplete published radial velocity
data  we find  that there  are at  least 14  radial-velocity confirmed
companions.  As probes of the  gravitational field or as indicators of
the  average number of  companions per  elliptical galaxy  the present
study  indicates that there  are at  least as  many companions  in the
vicinity of ellipticals (0.4 to 1.0 companions per galaxy in the inner
75~kpc,  and at  least  4  companions per  galaxy  in the  surrounding
300~kpc  sphere) as there  are around  spiral galaxies  (if Holmberg's
statistical numbers, giving 1.1 companions/spiral within 75~kpc, or if
Zaritsky,  Smith Frenk  \& White's  1993 value  of 1.5  companions per
late-type spiral are used as fiducial.) 

And finally, using the physical companions with published radial
velocities and differencing them against the redshift of the central
elliptical we can look at the ensemble fall-off of the one-dimensional
velocity dispersion as a function of radius. Figure 11 shows the
data. For the total sample there appears to be no significant trend,
with the velocity dispersion holding at a level of about 350~km/sec
over the entire range of radius. However, we note that in the
subsample expunging the galaxies in wide groups and clusters (numbers
shown in square brackets) there is the suggestion of a fall-off in the
observed velocity dispersion as a function of projected radial
distance (starting from a high of $\sim$370~km/s in the second bin and
dropping to 250-300~km/s in the outer bins). 

The above result is in general agreement with the larger study of
Praha et al. (2003) which motivated Figure 12. There we have plotted
the velocity dispersion as a function of radial separation binned so
as to more closely balance the number of companions in each of the
outer five bins. While a flat velocity-dispersion profile is
consistent with the data and their error bars over the entire 300~kpc
range, the dotted line is clearly a better fit to the data defining
the outermost 5 bins (120 to 300~kpc). This line is a Kelperian
fall-off scaled to a velocity dispersion of 250~km/s at 300~kpc. The
global significance of this fit and its consistency with the apparent
flattening of the density profile for the companion galaxies awaits
better statistics on a larger sample of galaxies.

In sum, we have demonstrated that the NASA/IPAC Extraglactic Database
is a useful resource for investigating the environments of galaxies as
multi-wavelength surveys continue to detect new galaxies.  As a
consequence, this updated catalog gives a much more robust measure of
the phase-space density of galaxies in small targeted volumes,
compared with earlier published visual estimates.  We have used this
technique to upgrade and update the results of BS77 with respect to
physical companions around ``isolated'' elliptical galaxies.  This new
analysis clearly shows the limitations of the BS77 approach.  In
particular, the visual approach to galaxy detection adopted by BS77
did not find all the surrounding galaxies, and was not able to
identify individual physical companions.  Furthermore, their early
understanding of large scale structure compromised even their
statistical comparisons.  Using a better defined sample and the now
available redshift information, our new analysis has shown that most
ellipticals do have at least three physical companions bound by a
relatively large velocity dispersion of $\pm$300-350~km/sec, inside a
radius of 300~kpc.  The ensemble properties of these physical
companions are consistent with halo masses of $3 \times 10^{12}
M_{\sun}$ out to an enclosing radius of 100~kpc. A fall-off in the
velocity dispersion with radial distances beyond 100~kpc may indicate
that the edge of dark matter halo surrounding these elliptical
galaxies may have detected.



\acknowledgments

This  research was  exclusively undertaken  using the  tools  and data
provided by  the 19 June 2003  release of the WEB-based  version of the
NASA/IPAC Extragalactic  Database (NED), which is operated  by the Jet
Propulsion  Laboratory,  California  Institute  of  Technology,  under
contract with the National Aeronautics and Space Administration.



\appendix

\section{Appendix I: Comments on Individual Systems}

\medskip
\centerline{\bf NGC 0584}

NGC  0584 has  one certain  physical companion,  NGC 0580,  within the
original 75~kpc search radius (Figure 1).  BS77 noted a second optical
companion.   Visual  inspection  strongly  suggests that  the  object,
2MASXi J0130541-064941 (diametrically  opposite to NGC~0580 across the
central  E galaxy) is  also a  physical companion  based on  its size,
proximity  and morphology.   No  redshift yet  exists  for this  2MASS
object, but follow-up studies of it would be useful.

Although  NGC  0584 is  the  dominant  member  (in linear  extent  and
luminosity) of  a small chain of  galaxies, it is  interesting to note
that  all  of the  members  with  measured  radial velocities  have  a
positive  redshift  with  respect  to NGC~0584,  suggesting  that  the
elliptical is not the center of  gravity of this grouping and that the
velocities are  probing a wider  potential. Similar comments  apply to
several  other  groupings  in   this  paper:  All  of  the  companions
associated with NGC~1052 have  negative radial velocities with respect
to  the elliptical  under study.   NGC~4697 has  five  companions, and
NGC~3640 has 8 companions, and NGC~5322 has 4 companions, all of which
have  positive  radial velocities  with  respect  to their  associated
elliptical.

\medskip
\centerline{\bf NGC 0720}

This galaxy and  its retinue of companions was  studied extensively by
Dressler,  Schechter \&  Rose  (1986).  Their  companion  No. 14  (PGC
006960),  as well  as  one  of Vader  \&  Charboyer's (1994)  objects,
[VC94]~015113-1403.0   are   good   candidates  for   being   physical
companions.   Indeed, two  other galaxies,  KUG 0147-138  ($\Delta$V =
--556~km/sec) and  DDO 015 ($\Delta$V  = +35~km/sec), are  also likely
physical companions,  albeit somewhat further afield  than our 300~kpc
radial  cut-off,   being  found  50~arcmin   (345~kpc)  and  73~arcmin
(505~kpc) away from NGC~0720, respectively.

\medskip
\centerline{\bf NGC 1052}

In  addition to the  two radial-velocity  confirmed objects,  a visual
inspection of the 75-kpc field surrounding NGC~1052 (Figure 2) reveals
two  other  highly  likely  companions  which presently  do  not  have
confirming  radial  velocity  data.   They are  [KKS2000]~04,  a  fine
low-surface-brightness dwarf spheroidal, to the south-east, and 2MASXi
J0241351-081024 to  the north-east.  However,  the grand-design spiral
galaxy  NGC~1042   is  only   15~arcmin  away,  suggesting   that  the
elliptical,  NGC~1052, is  by no  means  isolated, but  may share  the
potential with at  least two other sizeable galaxies,  the other being
NGC~1035.  This  mixed-morphology triplet has  already been cataloged,
and is known as KTS~018.

We  also  suggest  that  [VC94]  023858-0820.4  (which  has  a  radial
velocity, but at its published position has no obvious identification)
should be identified with 2MASXi J0241351-081024.

\medskip
\centerline{\bf NGC 4589}

There is a dwarf galaxy,  [HS98]~162, about 5~arcmin to the north-west
(Figure 7).  It is a prime  candidate for being  a physical companion,
and for a follow-up radial velocity measurement.

\medskip
\centerline{\bf NGC 5812}

A visual inspection of the  POSS image surrounding NGC~5813 (Figure 8)
reveals  two  prime candidates  for  physical  companions (and  radial
velocity followup) that have  gone unlisted in previous surveys. These
are NGC~5812:[MFB03]~1 a nucleated (dNE) dwarf at RA(2000) = 15:00:55,
DEC(2000) = -07:24:59, and  NGC~5812:[MFB03]~2 at RA(2000) = 15:00:48,
DEC(2000) = -07:27:42.

\section{Appendix II: Interlopers}

Praha {\it et al.} (2003) have suggested that interlopers are a
significant source of bias based on their study of a much larger
sample (2500 sq. degrees) of the sky. To assess the likelihood of
interlopers in our specific sample we undertook the following simple
test. For each galaxy we re-interogated NED using the same search
radius and the same velocity range but off-setting the search center
by 5 degrees to the north or south of the original target. Again we
dropped Fornax-cluster galaxies from the test.

With the exception of NGC~4494, we found single radial-velocity
selected ``interlopers'' in only 6 cases (NGC~3078, 3613, 3962, 4697,
5061 and 5813), two interlopers in the test of NGC~4742, and none in
the remaining 21 galaxies studied.  Some of these ``interlopers'' (NGC
3631 and ESO 566-G30) are galaxies comparable in size to our central
ellipticals, and as such they would not have been counted as
companions, or would have eliminated the original elliptical from the
``isolated'' category had they been accidentally in the field of
view. For NGC~4494 we hit on a background grouping of 4 galaxies
(reasonably associated with NGC~4250) at a redshift systematically
higher than NGC~4494 by about 750~km/s.

From this test we conclude that our sample is probably contaminated at
the 5-10\% level by Hubble-flow interlopers in the general field, but
that is still well within the counting statistics of this modest
survey.

\medskip




\clearpage


\begin{figure}
\caption{(omitted from astro-ph version due to size limitations)The
75~kpc region around NGC~0584. The field of view shown here, and in
the following eight figures, is set to the same physical size of
75~kpc on a side.  The two radial-velocity-confirmed galaxies,
2MASXi~J0130541-064941 and NGC~0586 are labeled.
\label{fig1}}
\end{figure}

\clearpage 

\begin{figure}
\caption{(omitted from astro-ph version due to size limitations)The
75~kpc region around NGC~1052.  Only radial-velocity-confirmed
companions are labeled, with the exception of the giant
low-surface-brightness object [KKS2000]~04, and the 2MASS object
which, as noted in the Appendix, is thought by us to be the same as
the [VC94] galaxy which has a published radial velocity (but no
obvious optical counterpart on the POSS).  Note the large spiral
galaxy, NGC~1042, in the lower right corner of the frame, is just
outside the 75~kpc radius adopted by Bothun \& Sullivan (1977) for
their ``isolation'' criterion.
\label{fig2}}
\end{figure}

\clearpage 

\begin{figure}
\caption{(omitted from astro-ph version due to size limitations) The
75~kpc region around NGC~3348.  The single radial-velocity-confirmed
galaxy MCG +12-10-079 is labeled.
\label{fig3}}
\end{figure}

\clearpage 

\begin{figure}
\caption{(omitted from astro-ph version due to size limitations) The
75~kpc region around NGC~3640.  The single radial-velocity-confirmed
galaxy NGC~3641 is labeled.
\label{fig4}}
\end{figure}

\clearpage 

\begin{figure}
\caption{(omitted from astro-ph version due to size limitations) The
75~kpc region around NGC~3818.  The single radial-velocity-confirmed
galaxy 2dFGPS~N113Z117 is labeled.
\label{fig5}}
\end{figure}

\clearpage 

\begin{figure}
\caption{(omitted from astro-ph version due to size limitations) The
75~kpc region around NGC~4125.  The single radial-velocity-confirmed
galaxy NGC~4121 is labeled.
\label{fig6}}
\end{figure}

\clearpage 

\begin{figure}
\caption{(omitted from astro-ph version due to size limitations) The
75~kpc region around NGC~4589.  The single radial-velocity-confirmed
galaxy NGC~4572 is labeled.
\label{fig7}}
\end{figure}

\clearpage 

\begin{figure}
\caption{(omitted from astro-ph version due to size limitations) The
75~kpc region around NGC~5812.  The single radial-velocity-confirmed
galaxy IC~1084 is labeled.
\label{fig8}}
\end{figure}

\clearpage 

\begin{figure}
\caption{(omitted from astro-ph version due to size limitations) The
75~kpc region around NGC~5831.  The two radial-velocity-confirmed
galaxies, NGC~5846:[ZM98]~0028 and NPM1G~+01.0437 are labeled.
\label{fig9}}
\end{figure}

\clearpage 

\begin{figure}
\plotfiddle{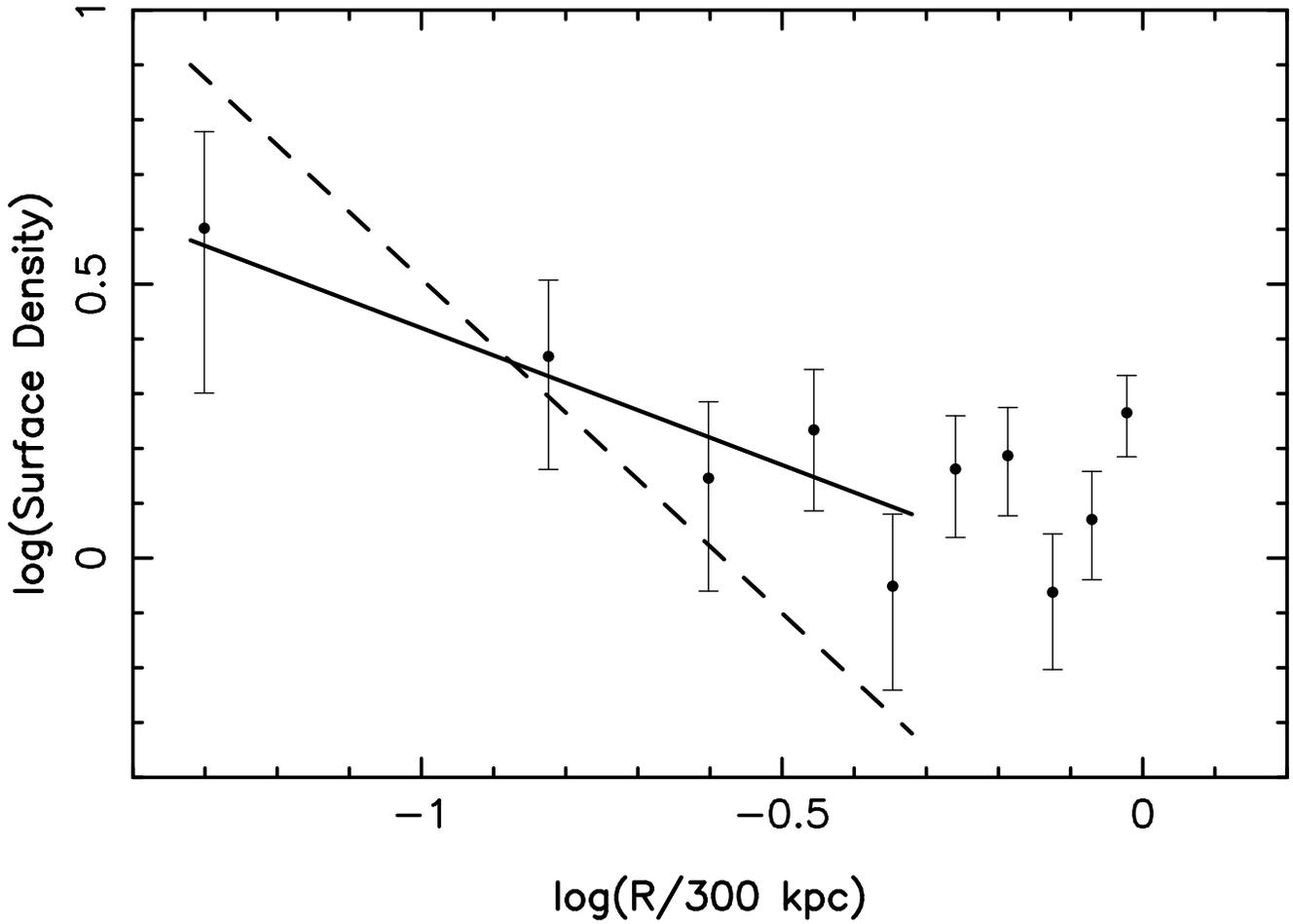}{1.0in}{-90.}{400.}{500.}{-25}{2}
\caption{ The  radial density distribution for the  entire ensemble of
radial-velocity-confirmed companions  to 34 isolated  ellipticals. The
solid line  has a  slope $\alpha =  0.5$ ,  while the dashed  line has
slope $\alpha  = -1.22$,  as suggested by  Vader \& Sandage  (1991) in
their study of companions to  RSA galaxies.  A fall-off is apparent no
further than 100~kpc after which  the density of companions appears to
be relatively constant.
\clearpage
\label{fig10}}
\end{figure}

\clearpage 

\begin{figure}
\plotfiddle{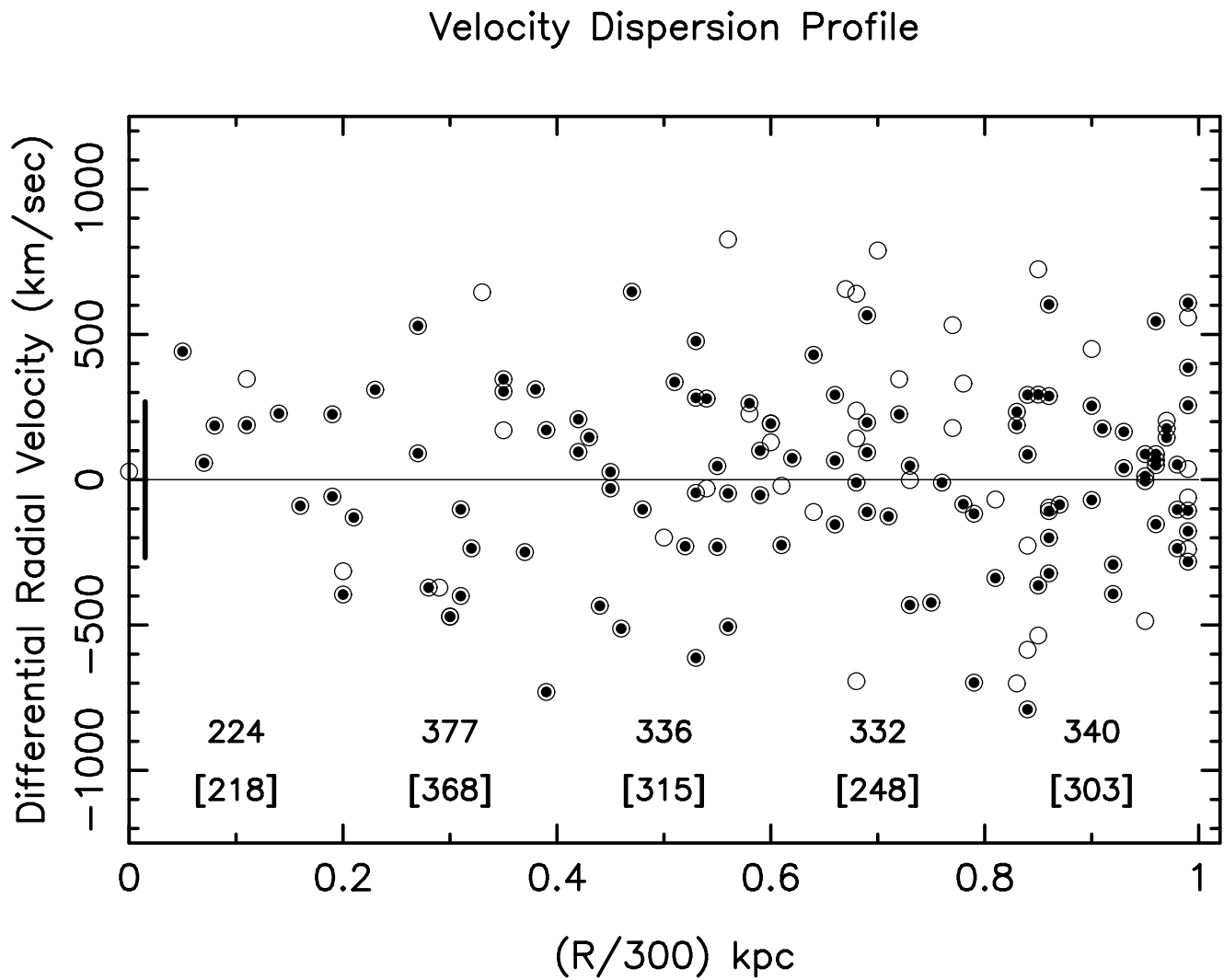}{1.0in}{-90.}{400.}{500.}{-2}{25}
\caption{ Individual velocity differences as a function of distance
from the central elliptical galaxy.  The open circles are the 48
radial-velocity companions along the line of sight to the five Fornax
cluster ellipticals.  Velocity dispersion for the entire sample,
(calculated in the 5 bins, each of width 60~kpc) are shown in the
lower part of the figure.  The generally smaller numerical values
below them (in square brackets) are the dispersions calculated for the
non-Fornax sample (plotted as circled dots).  No statistically
significant change in velocity dispersion with radial separation from
the central elliptical is apparent for the larger sample; however,
there is an indication of a decline in velocity dispersion with radius
for the last four bins in the non-Fornax sample (see Figure 12).
\label{fig11}}
\end{figure}
\clearpage

\begin{figure}
\plotfiddle{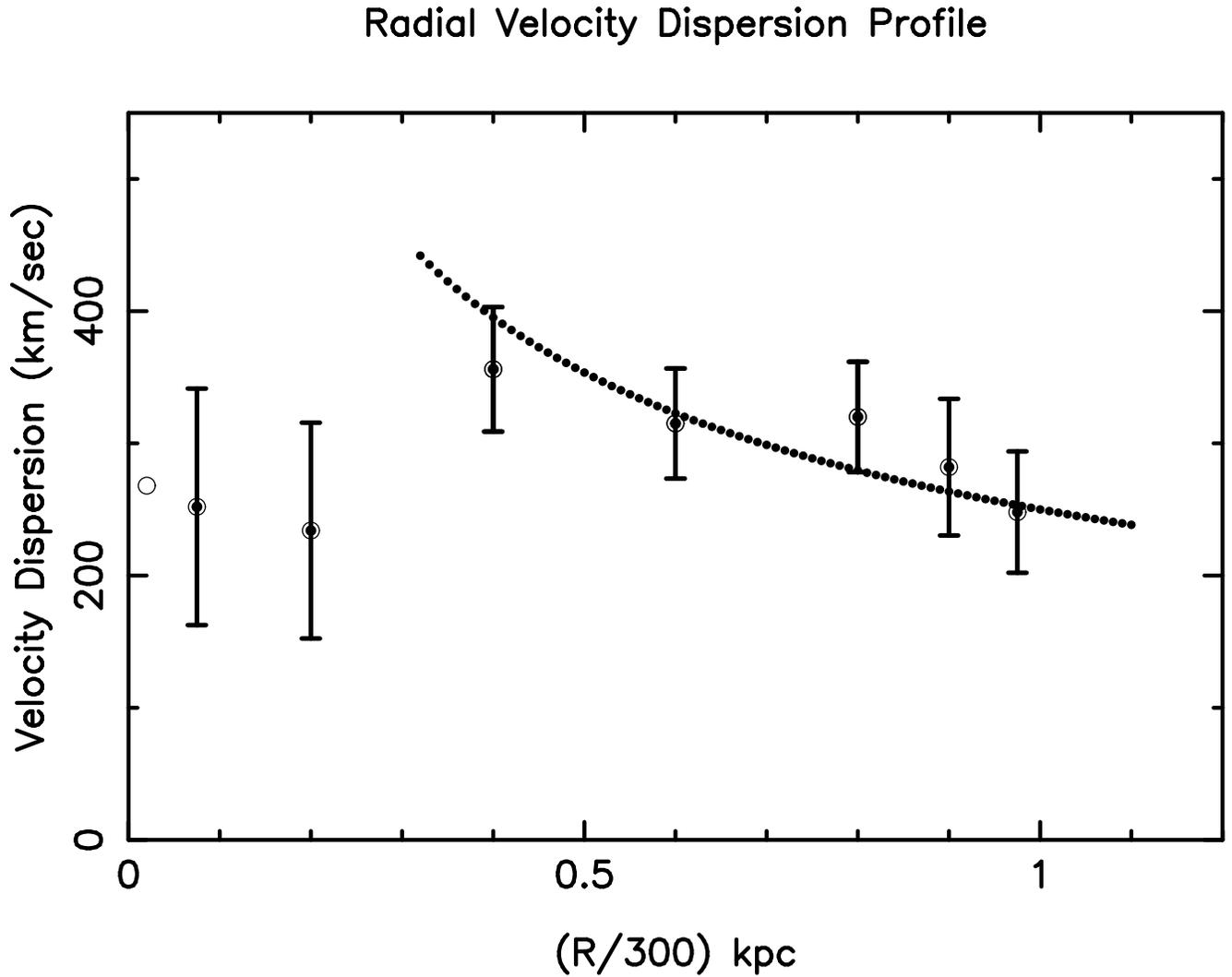}{1.0in}{-90.}{400.}{500.}{-2}{20}
\caption{ The run of velocity dispersion as a function of distance
from the central elliptical galaxy for the non-Fornax sample. For the
outer five points the binning was chosen so as to equalize the error
bars by placing approximately the same number of companions in each
bin.  The apparent fall-off in velocity dispersion with radius begins
at approximately 120~kpc (R/300 = 0.4) and continues to the last
observed point at 300~kpc. The Keplerian fall-off having a velocity
dispersion of 250~km/sec at 300~kpc as shown by the dotted line is a
reasonable fit to the data. The open circle near R = 0.0 is the
average of the central velocity dispersions of the parent elliptical
galaxies
\label{fig12}}
\end{figure}

\clearpage




\setcounter{page}{25}

\clearpage

\begin{table}
\tablenum{1}
\caption{Previous Results on Companion Frequencies and Radial Fall-Off}
\begin{tabular}{lccccc}

\hline

\hline

& Author(s) &Central Galaxy&Average &Search & Fall-off \\

& Reference &Type &No. Comps.&Radius& Index\tablenotemark{*}\\

\hline

& Holmberg (1969) & S/S0 & 1.1 $\pm$ 0.1 & 75 kpc& . . .\\
& Bothun \& Sullivan (1972) & E & 0.1 $\pm$ 0.4 & 75 kpc& . . .\\
& Lake \& Tremaine (1980) & S/S0 & . . .  & 75 kpc& --0.5 $\pm$ 0.2\\
& Zaritsky et al. (1993) & S & 1.5 $\pm$ 0.4 & 75 kpc& --1.0 $\pm$ 0.2\\
& Vader \& Sandage (1991) & S/S0 & 1.0 $\pm$ 0.1 & 16-270 kpc& --1.22 $\pm$ 0.05\\
& McKay et al. (2002) & E/S/S0 & 1.0 & 260 kpc& . . .\\
& Brainerd \& Specian (2003) & E/S/S0 & 1.9 $\pm$ 0.1& 260 kpc& --1.1 $\pm$ 0.1\\
& Prada et al. (2003) & E/S/S0 & 0.36 $\pm$ 0.02& 500 kpc& . . .\\
& Madore, Freedman \& Bothun (this paper)  & E & 0.74 $\pm$ 0.95 & 75 kpc& . . .\\
& Madore, Freedman \& Bothun (this paper) & E & 4.3 $\pm$ 0.4  & 300 kpc&--0.5\\
& Garcia-Barret et al. (2003) & SB & 3-4  & 1000 kpc& . . .\\

\hline

\hline

\end{tabular}
\tablenotetext{*}{The fall-off index is given here for the surface density. Augment
the index by -1 for the volume-density fall-off. Mixed types of
indices are published in the original articles.}
\end{table}

\clearpage
\begin{table}
\tablenum{2}
\tablewidth{5.0truein}
\caption{Companion-Galaxy Counts within R = 75~Kpc of `Isolated' Ellipticals}
\begin{tabular}{ccrrrrrcrr}
\hline
\hline
& Name &N75 &[N]~&[S]~&Excess~~ &BS77& V[GSR](km/s)~&S75~~&L150\\ 
& (1) &(2)~ &(3) &(4) &(5)~~~~ &(6)~~~& (7)& (8)&(9)\\
\hline
& NGC~0584 & 4 & 1  & 1 & +3.0$\pm$2.2 & +1.5 &1,837&1&4\\
& NGC 0636 & 3 & 3  & 2 & +0.5$\pm$2.3 & --1.0 &1,933&1&1\\
& NGC 0720 & 1 & 0  & 6 &--2.0$\pm$2.0 & +0.0 &1,826&0&6\\
& NGC 0821 & 2 & 0  & 0 & +2.0$\pm$1.4 & --1.5 &1,799&0&0\\
& NGC 1052 & 4 & 3  & 8 &--1.5$\pm3.1$ & --0.5  &1,455&2&4\\
&&&&&&&\\
& NGC 1351 & 80 & 80 & 96 & --8.0$\pm$13.0 & +0.5  &1,593&0&8\\
& NGC 1395 & 39 & 60 & 42 &--12.0$\pm$9.5 & +0.0  &1,622&0&5\\
& NGC 1426 & 4 & 1 & 25 &   --9.0$\pm$4.1 & +1.5  &1,348&0&9\\
& NGC 1427 & 134 & 75 & 60 &+65.5$\pm$14.2  & --0.5  &1,263&2&28\\
& NGC 1439 & 2 & 3 & 6 &    --2.5$\pm$2.5 & +1.5  &1,575&0&2\\
&&&&&&&\\
& NGC 2768 & 5 & 0 & 1 & +4.5$\pm$2.3 & +0.0  &1,444&0&3\\
& NGC 2974 & 6 & 1 & 2 & +4.5$\pm$2.7 & +0.5  &1,908&0&0\\
& NGC 2986 & 3 & 0 & 0 & +3.0$\pm$1.7 & +1.0  &2,097&0&0\\
& NGC 3078 & 0 & 0 & 0 & +0.0$\pm$0.0 & --1.0  &2,283&0&3\\
& NGC 3348 & 2 & 1 & 6 &--1.5$\pm$2.3 & +0.0  &2,960&1&1\\
&&&&&&&\\
& NGC 3585 & 0 & 0 & 1 & --0.5$\pm$0.7 & +0.0  &1,206&0&2\\
& NGC 3610 & 1 & 0 & 3 & --0.5$\pm$1.6 & --0.5  &1,777&0&4\\
& NGC 3613 & 3 & 2 & 2 & +1.0$\pm$2.2 & --0.5  &2,066&0&9\\
& NGC 3640 & 15 & 15 & 50 & --17.5$\pm$6.9 & +0.5  &1,198&1&8\\
& NGC 3818 & 46 & 80 & 49 & --18.5$\pm$10.5 & +1.0  &1,566&1&3\\
&&&&&&&\\
& NGC 3904 & 1 & 0 & 1 & +0.5$\pm$1.2 & +0.5  &1,393&0&4\\
& NGC 3923 & 5 & 0 & 1 & +4.5$\pm$2.3 & +1.5  &1,557&0&5\\
& NGC 3962 & 3 & 3 & 4 & --0.5$\pm$2.5 & --0.5  &1,666&0&0\\
& NGC 4125 & 9 & 2 & 2 & +7.0$\pm$3.3 & +3.0  &1,469&1&2\\
& NGC 4494 & 20 & 18 & 24 & --1.0$\pm$6.4 & --0.5  &1,343&0&0\\
&&&&&&&\\
& NGC 4589 & 9 & 1 &3  & +7.0$\pm$3.3 & +0.5  &2,122&1&6\\
& NGC 4697 & 81 & 85 & 96 & --9.5$\pm$13.1 & --2.0  &1,141&0&6\\
& NGC 4742 & 6 & 9 & 10 & --3.5$\pm$3.9 & --3.5  &1,160&0&7\\
& NGC 5061 & 3 & 5 & 6 & --2.5$\pm$2.9 & --0.5  &2,525&0&0\\
& NGC 5322 & 4 & 3 & 6 & --0.5$\pm$2.9 & +0.5  &1,908&0&4\\
&&&&&&&\\
& NGC 5812 & 1 & 0 & 0 & +1.0$\pm$1.0 & +2.0  &1,911&1&3\\
& NGC 5813 & 3 & 0 & 0 & +3.0$\pm$1.7 & +1.0  &1,980&0&5\\
& NGC 5831 & 6 & 2 & 7 & +1.5$\pm$3.2 & --0.5  &1,665&2&11\\
& NGC 6958 & 24 & 29 & 25 &+9.5$\pm$13.1 & --2.0 &2,730&0&1\\

\hline

\end{tabular}
\end{table}

\clearpage
\begin{table}
\tablenum{3}
\caption{Velocity-Confirmed Companions within R = 300~kpc}
\begin{tabular}{llrr}
\hline

\hline

& {$\underline{\bf Central~E}$} (Central $\sigma$) &R~~~~~~&$\Delta$V(comp-E)\\
& Companion(s) &(arcmin)&(km/sec)\\

\hline

&  {$\underline{\bf NGC~0584}$ ($\sigma_c = \pm$225~km/sec)} &&    \\

& NGC 0586~* & 4.3~~ & +188~~ \\ 
& IC 0127 & 24.0~~ & +193~~ \\ 
& NGC 0596 & 24.7~~  & +74~~ \\ 
& NGC 0600 & 37.2~~ & +40~~ \\ 
& KDG 007 ** & 58.9~~ & --961~~ \\ 
&&&\\
&  {$\underline{\bf NGC~0636}$ ($\sigma_c = \pm$167~km/sec)} &&    \\

&{MCG --01--05--012~*} & 16.8~~ &+96~~\\
&&&\\
&  {$\underline{\bf NGC~0720}$ ($\sigma_c = \pm$237~km/sec)} &&    \\

&KUG 0150-138 &11.3~~& --371~~\\
&LEDA 087905 &14.8~~& --249~~\\
&LEDA 087906 &18.6~~& --512~~\\
&MCG -02-05-074 &27.7~~&+197~~\\
&LEDA 087900 &32.3~~& --338~~\\
&MCG -02-05-072 &34.4~~& --322~~\\
&KUG 0147-138 ** &50.2~~& --296~~\\
&&&\\
&{$\underline{\bf NGC~0821}$ ($\sigma_c = \pm$209~km/sec)}&&\\

&n/a&. . .~~&. . .~~\\
&&&\\
&{$\underline{\bf NGC 1052}$ ($\sigma_c = \pm$215~km/sec)}&&\\

&[VC94] 023858-0820.4~* &9.3~~&--58~~\\
&NGC 1047~*&10.2~~&--130~~\\
&NGC 1042&14.7~~&--102~~\\
&NGC 1035&24.8~~&--229~~\\
&&&\\
&{$\underline{\bf NGC 1351}$ ($\sigma_c = \pm$147~km/sec)}&&\\

&FCC 100&19.1~~&+146~~\\
&NGC 1351A&29.2~~&--154~~\\
&MCG -06-08-025&30.4~~&--111~~\\
&ESO 358-015&31.3~~&--126~~\\
&FCC 085&41.7~~&--5~~\\
&FCCB 905&44.0~~&--236~~\\
&ESO 358-006&45.0~~&--177~~\\
&LSBG F358-61&45.9~~&+609~~\\
&NGC 1373 **&57.4~~&--280~~\\
&ESO 358-019 **&58.9~~&--260~~\\

\hline

\end{tabular}
\end{table}

\clearpage
\begin{table}
\tablenum{3}
\caption{Velocity-Confirmed Companions within R = 300~kpc}
\begin{tabular}{llrr}
\hline

\hline

& {$\underline{\bf Central~E}$} (Central $\sigma$) &R~~~~~~&$\Delta$V(comp-E)\\
& Companion(s) &(arcmin)&(km/sec)\\

\hline

&{$\underline{\bf NGC 1395}$ ($\sigma_c = \pm$248~km/sec)}&&\\

&ESO 482-017&12.8~~&--371~~\\
&NGC 1401&21.8~~&--199~~\\
&ESO 482-018&23.7~~&--30~~\\
&ESO 482-031&37.8~~&--96~~\\
&NGC 1416&39.8~~&+450~~\\
&NGC 1415 **&43.8~~&--132~~\\
&2MASX J03412708-2228220 **&52.7~~&+553~~\\
&ESO 548-070 ** &53.8~~&--295~~\\
&ESO 482-036 **&55.3~~&--150~~\\
&&&\\
&  {$\underline{\bf NGC~1427}$ ($\sigma_c = \pm$170~km/sec)} &&    \\

&CGF 10-18~*  &0.1~~ &+28~~ \\ 
&FCCB 1554~*  &4.9~~ &+347~~ \\ 
&FCC 274  &8.9~~ &--315~~ \\ 
&NGC 1428  &14.3~~ &+252~~ \\ 
&FCC 264  &15.3~~ &+645~~ \\ 
&FCC 266  &15.4~~ &+170~~ \\ 
&LSBG F358-37  &24.6~~ &+657~~ \\ 
&LSBG F358-34  &25.5~~ &--291~~ \\ 
&LSBG F358-33  &26.4~~ &+129~~ \\ 
&FCC 252  &28.0~~ &--111~~ \\ 
&APMUKS(BJ) B033815.38-354624.4 &29.5~~ &+656~~ \\ 
&FCC 296  &29.7~~ &--693~~ \\ 
&NGC 1427A  &29.8~~ &+640~~ \\ 
&FCSS J034007.2-353705  &30.1~~ &+670~~ \\ 
&FCC 245  &30.9~~ &+789~~ \\ 
&ESO 358-051  &31.7~~ &+346~~ \\ 
&NGC 1436  &31.9~~ &--1~~ \\ 
&FCSS J033935.9-352824  &33.9~~ &+532~~ \\ 
&FCC 298  &34.3~~ &+331~~ \\ 
&FCSS J033952.5-350424  &35.6~~ &--68~~ \\ 
&LSBG F358-36  &36.6~~ &--701~~ \\ 
&ESO 358-060  &36.8~~ &--585~~ \\ 
&AM 0337-353  &37.9~~ &--536~~ \\ 
&AM 0337-355  &41.7~~ &--486~~ \\ 
&FCSS J033854.1-353333  &43.0~~ &+203~~ \\ 
&NGC 1404  &43.9~~ &+559~~ \\ 
&NGC 1399  &47.0~~ &+37~~ \\ 
&FCC 230  &47.4~~ &--239~~ \\ 

\hline

\end{tabular}
\end{table}

\clearpage
\begin{table}
\tablenum{3}
\caption{Velocity-Confirmed Companions within R = 300~Kpc}
\begin{tabular}{llrr}
\hline

\hline

& {$\underline{\bf Central~E}$} (Central $\sigma$) &R~~~~~~&$\Delta$V(comp-E)\\
& Companion(s) &(arcmin)&(km/sec)\\

\hline

&  {$\underline{\bf NGC~1426}$ ($\sigma_c = \pm$153~km/sec)} &&    \\

&2MASXi J0341270-222822  &29.0~~ &+827~~ \\ 
&NGC 1439  &30.2~~ &+227~~ \\ 
&NGC 1422  &31.4~~ &+194~~ \\ 
&ESO 548-070  &31.5~~ &--21~~ \\ 
&NGC 1414  &35.2~~ &+238~~ \\ 
&NGC 1415  &37.7~~ &+142~~ \\ 
&ESO 482-036  &39.3~~ &+124~~ \\ 
&ESO 482-031  &43.9~~ &+178~~ \\ 
&NGC 1416  &44.1~~ &+724~~ \\ 
&ESO 549-006 ** &53.5~~ &+166~~ \\ 
&&&\\
&  {$\underline{\bf NGC~1439}$ ($\sigma_c = \pm$159~km/sec)} &&    \\

&NGC 1426  &30.2~~ &--227~~ \\ 
&ESO 549-006  &44.2~~ &--61~~ \\ 
&NGC 1422 ** &48.3~~ &--33~~ \\ 
&ESO 549-018 ** &54.4~~ &--83~~ \\ 
&NGC 0141 **  &55.5~~ &+11~~ \\ 
&2MASX J03412708-2228220 ** &57.5~~ &+600~~ \\ 
&&&\\
&  {$\underline{\bf NGC~2768}$ ($\sigma = \pm$194~km/sec)} &&    \\

&CGCG 288-027 NED01  &16.5~~ &--236~~ \\ 
&NGC 2742  &40.2~~ &--84~~ \\ 
&NGC 2726  &50.5~~ &+145~~ \\ 
&&&\\
&  {$\underline{\bf NGC~2974}$ ($\sigma_c = \pm$220~km/sec)} &&    \\

&PGC 027612 ** &33.7~~ &--559~~ \\ 
&&&\\
&  {$\underline{\bf NGC~2986}$ ($\sigma_c = \pm$268~km/sec)} &&    \\

&NGC 2983 ** &48.7~~ &--287~~ \\ 
&&&\\
&  {$\underline{\bf NGC~3078}$ ($\sigma_c = \pm$237~km/sec)} &&    \\

&ESO 499-032  &14.8~~ &--45~~ \\ 
&NGC 3048  &15.3~~ &+47~~ \\ 
&ESO 499-022  &32.9~~ &--282~~ \\ 
&&&\\
&  {$\underline{\bf NGC~3348}$ ($\sigma_c = \pm$237~km/sec)} &&    \\

&NGC 3364  &25.6~~ &--106~~ \\ 

\hline

\end{tabular}
\end{table}

\clearpage
\begin{table}
\tablenum{3}
\caption{Velocity-Confirmed Companions within R = 300~Kpc}
\begin{tabular}{llrr}
\hline

\hline

& {$\underline{\bf Central~E}$} (Central $\sigma$) &R~~~~~~&$\Delta$V(comp-E)\\
& Companion(s) &(arcmin)&(km/sec)\\

\hline

&  {$\underline{\bf NGC~3585}$ ($\sigma_c = \pm$216~km/sec)} &&    \\

&2MASX J11142003-2654462 &16.9~~ &+304~~ \\ 
&ESO 503-007 &40.0~~ &+188~~ \\ 
&&&\\
&  {$\underline{\bf NGC~3610}$ ($\sigma_c = \pm$167~km/sec)} &&    \\

&SBS 1114+587 &19.2~~&--102~~ \\ 
&UGC 06304 &26.5~~&+66~~ \\ 
&SBS 1115+585 &28.8~~&+225~~ \\ 
&NGC 3642 &34.6~~&--108~~ \\ 
&SBS 1119+586 ** &42.9~~&--113~~ \\ 
&UGC 06412 **&46.6~~&+310~~ \\ 
&NGC 3613 **&47.2~~&+291~~ \\ 
&SBS 1119+583 **&55.2~~&--73~~ \\ 
&&&\\
&  {$\underline{\bf NGC~3613}$ ($\sigma_c = \pm$213~km/sec)} &&    \\

&NGC 3619&15.7~~&--434~~\\
&SBS 1115+585 &18.5~~&+336~~\\ 
&NGC 3625&20.1~~&--47~~\\
&UGC 06344&21.1~~&--53~~\\
&UGC 06304&22.0~~&--225~~\\
&SBS 1119+583&30.5~~&--364~~\\
&SBS 1118+578A&32.7~~&+176~~\\
&SBS 1114+587&33.2~~&--393~~\\
&SBS 1118+578B&34.8~~&+176~~\\
&SBS 1119+583 **&37.5~~&+0~~\\
&&&\\
&  {$\underline{\bf NGC~3640}$ ($\sigma_c = \pm$184~km/sec)} &&    \\

&NGC 3641~* &2.5~~&+441~~\\
&NGC 3643&14.0~~&+529~~\\
&NGC 3630&20.5~~&+171~~\\
&UM 442 &44.2~~&+293~~\\
&UGC 06345&44.5~~&+288~~\\
&NGC 3664A&49.5~~&+12~~\\
&NGC 3664&49.8~~&+68~~\\
&UGC 06417&49.9~~&+50~~\\
&&&\\
&  {$\underline{\bf NGC~3818}$ ($\sigma_c = \pm$198~km/sec)} &&    \\

&2dFGPS N113Z117~* &4.0~~&+186~~\\
&UGCA 242 &21.4~~&+27~~\\
&LCRS B113807.1-053433 &26.4~~&--231~~\\

\hline

\end{tabular}
\end{table}

\clearpage
\begin{table}
\tablenum{3}
\caption{Velocity-Confirmed Companions within R = 300~Kpc}
\begin{tabular}{llrr}
\hline

\hline

& {$\underline{\bf Central~E}$} (Central $\sigma$) &R~~~~~~&$\Delta$V(comp-E)\\
& Companion(s) &(arcmin)&(km/sec)\\

\hline

&  {$\underline{\bf NGC~3904}$ ($\sigma_c = \pm$200~km/sec)} &&    \\

&NGC 3923 &36.9~~ &+292~~ \\ 
&ESO 440-014 &37.9~~ &+603~~ \\ 
&2MASXi J1151382-284734 &42.9~~ &--103~~ \\ 
&ESO 440-023 &44.7~~ &+386~~ \\ 
&ESO 440-018 ** &58.4~~ &+90~~ \\ 
&UGCA 247 ** &59.2~~ &+359~~ \\ 
&&&\\
&  {$\underline{\bf NGC~3923}$ ($\sigma_c = \pm$241~km/sec)} &&    \\

&2MASXi J1151382-284734 &8.1~~ &--395~~ \\ 
&ESO 440-014 &15.2~~ &+311~~ \\ 
&ESO 440-023 &27.6~~ &+94~~ \\ 
&UGCA 250 &34.6~~ &--86~~ \\ 
&NGC 3904 &36.9~~ &--292~~ \\ 
&FLASH J115153.41-281046.5 ** &39.3~~ &--237~~ \\ 
&UGCA 247 **&42.8~~ &+194~~ \\ 
&FLASH J114755.66-281156.2 ** &54.8~~ &--323~~ \\ 
&ESO 440-030 ** &58.0~~ &+82~~ \\ 
&&&\\
&  {$\underline{\bf NGC~3962}$ ($\sigma_c = \pm$225~km/sec)} &&    \\

&n/a &. . . ~~ &. . .~~ \\ 
&&&\\
&  {$\underline{\bf NGC~4125}$ ($\sigma_c = \pm$233~km/sec)} &&    \\

&NGC 4121~* &3.7~~ &+58~~ \\ 
&NGC 4081 &49.6~~ &+88~~ \\ 
&UGC 07020A **&59.3~~ &+158~~ \\ 
&&&\\
&  {$\underline{\bf NGC~4494}$ ($\sigma_c = \pm$155~km/sec)} &&    \\

&NGC 4562 **&56.6~~ &+2~~ \\ 
&&&\\
&  {$\underline{\bf NGC~4589}$ ($\sigma_c = \pm$225~km/sec)} &&    \\

&NGC 4572~* &7.5~~ &+225~~ \\ 
&NGC 4648 &22.3~~ &--506~~ \\ 
&UGC 07844 &31.7~~ &--117~~ \\ 
&UGC 07767 &32.0~~ &--698~~ \\ 
&UGC 07745 &33.5~~ &--790~~ \\ 
&UGC 07908 **&43.3~~ &--320~~ \\ 

\hline

\end{tabular}
\end{table}

\clearpage
\begin{table}
\tablenum{3}
\caption{Velocity-Confirmed Companions within R = 300~Kpc}
\begin{tabular}{llrr}
\hline

\hline

& {$\underline{\bf Central~E}$} (Central $\sigma$) &R~~~~~~&$\Delta$V(comp-E)\\
& Companion(s) &(arcmin)&(km/sec)\\

\hline

&  {$\underline{\bf NGC~4697}$ ($\sigma_c = \pm$174~km/sec)} &&    \\

&MCG -01-33-007 &14.9~~ &+91~~ \\ 
&DDO 148 &32.8~~ &+100~~ \\ 
&2MASXi J1250191-052146 &36.7~~ &+292~~ \\ 
&DDO 146 &46.3~~ &+234~~ \\ 
&NGC 4731 &50.6~~ &+254~~ \\ 
&NGC 4731A &59.9~~ &+256~~ \\ 
&&&\\
&  {$\underline{\bf NGC~4742}$ ($\sigma_c = \pm$102~km/sec)} &&    \\

&NGC 4757 &17.6~~ &--400~~ \\ 
&NGC 4781 &38.6~~ &--10~~ \\ 
&MCG -06-33-015 &40.8~~ &+48~~ \\ 
&[KEB] J124959.6-093036 &42.1~~ &--423~~ \\ 
&NGC 4784 &42.6~~ &--10~~ \\ 
&NGC 4790 &46.9~~ &+87~~ \\ 
&UGC 308 &55.0~~ &52~~ \\ 
&&&\\
&  {$\underline{\bf NGC~5061}$ ($\sigma_c = \pm$194~km/sec)} &&    \\

&ESO 508-033 ** &28.1~~ &+669~~ \\ 
&ESO 508-039 ** &36.0~~ &--586~~ \\ 
&IC 0879 ** &41.4~~ &--692~~ \\ 
&NGC 5078 ** &41.6~~ &--493~~ \\ 
&IC 0874 ** &49.1~~ &+442~~ \\ 
&ESO 508-051 ** &55.5~~ &--528~~ \\ 
&&&\\
&  {$\underline{\bf NGC~5322}$ ($\sigma_c = \pm$234~km/sec)} &&    \\

&UGC 08716 &21.0~~ &+282~~ \\ 
&UGC 08741 &21.4~~ &+279~~ \\ 
&UGC 08714 &23.2~~ &+263~~ \\ 
&NGC 5342 &25.5~~ &+430~~ \\ 
&NGC 5308 ** &49.8~~ &+111~~ \\ 
&NGC 5379 ** &54.5~~ &--323~~ \\ 
&NGC 5389 ** &58.1~~ &--88~~ \\ 
&UGC 08859 ** &58.7~~ &--320~~ \\ 
&&&\\
&  {$\underline{\bf NGC~5812}$ ($\sigma_c = \pm$213~km/sec)} &&    \\

&IC 1084~* &5.0~~ &+228~~ \\ 
&LCRS B145849.9-064423 &32.3~~ &--70~~ \\ 
&MCG -01-38-014 &33.6~~ &+165~~ \\ 
&MCG -01-38-020 **&51.0~~ &--75~~ \\ 
&&&\\
\hline

\end{tabular}
\end{table}

\clearpage
\begin{table}
\tablenum{3}
\caption{Velocity-Confirmed Companions within R = 300~Kpc}
\begin{tabular}{llrr}
\hline

\hline

& {$\underline{\bf Central~E}$} (Central $\sigma$) &R~~~~~~&$\Delta$V(comp-E)\\
& Companion(s) &(arcmin)&(km/sec)\\

\hline

&  {$\underline{\bf NGC~5813}$ ($\sigma = \pm$230~km/sec)} &&    \\

&NGC 5811 NED02 &11.9~~ &--471~~ \\ 
&NGC 5811 NED01 &12.1~~ &--471~~ \\ 
&UGC 09661 &15.5~~ &--730~~ \\ 
&NGC 5806 &21.0~~ &--613~~ \\ 
&CGCG 020-042 &38.4~~ &--153~~ \\ 
&CGCG 020-039 ** &40.5~~ &--162~~ \\ 
&NGC 5846:[ZM98] 0021 ** &45.5~~ &--108~~ \\ 
&NGC 5831 **&52.6~~ &--316~~ \\ 
&2MASX J15035028+0107366 **&52.6~~ &--406~~ \\ 
&NGC 5846:[ZM98] 0028 ** &58.3~~ &--6~~ \\ 
&NGC 5846:[ZM98] 0017 ** &58.9~~ &+30~~ \\ 
&MRK 1390 ** &59.7~~ &+100~~ \\ 

\hline

\end{tabular}
\end{table}

\clearpage
\begin{table}
\tablenum{3}
\caption{Velocity-Confirmed Companions within R = 300~Kpc}
\begin{tabular}{llrr}
\hline

\hline

& {$\underline{\bf Central~E}$} (Central $\sigma$) &R~~~~~~&$\Delta$V(comp-E)\\
& Companion(s) &(arcmin)&(km/sec)\\

\hline

&  {$\underline{\bf NGC~5831}$ ($\sigma = \pm$168~km/sec)} &&    \\

&NPM1G +01.0437~\tablenotemark{*} &7.0~~ &--90~~ \\ 
&NGC 5846:[ZM98] 0028~\tablenotemark{*} &9.9~~ &+310~~ \\ 
&NGC 5846:[ZM98] 0017 &15.2~~ &+346~~ \\ 
&NGC 5846:[ZM98] 0021 &18.3~~ &+208~~ \\ 
&NGC 5846:[ZM98] 0033 &20.8~~ &+647~~ \\ 
&NGC 5846:[ZM98] 0039 &23.2~~ &+477~~ \\ 
&NGC 5846:[ZM98] 0062  &30.5~~ &+566~~ \\ 
&NGC 5839 &32.0~~ &--431~~ \\ 
&NGC 5845 &37.8~~ &--200~~ \\ 
&NGC 5846A &42.1~~ &+545~~ \\ 
&NGC 5846 &42.4~~ &+88~~ \\ 
&NGC 5846:[ZM98] 0049\tablenotemark{**} &46.9~~ &+721~~ \\ 
&KKR 15 &47.5~~ &--68~~ \\ 
&UGC 09661\tablenotemark{**} &48.4~~ &--414~~ \\ 
&NGC 5846:[ZM98] 0026\tablenotemark{**} &48.7~~ &+173~~ \\ 
&NGC 5850\tablenotemark{**} &49.2~~ &+870~~ \\ 
&NGC 5813\tablenotemark{**} &52.6~~ &+316~~ \\ 
&MRK 1390\tablenotemark{**} &55.4~~ &+379~~ \\ 
&NGC 5838\tablenotemark{**} &56.4~~ &--296~~ \\ 
&NGC 5841\tablenotemark{**} &59.9~~ &--387~~ \\ 
&&&\\
&  {$\underline{\bf NGC~6958}$ ($\sigma = \pm$223~km/sec)} &&    \\

&APMBGC 341+016-108 &12.7~~ &-30~~ \\ 

\hline
\hline

\end{tabular}
\tablenotetext{*}{Indicates an object that is within the R = 75~kpc sample.}
\tablenotetext{**}{Indicates an object that is  within a
one-degree search radius, but beyond the 300~kpc cut-off radius analyzed in this paper}
\end{table}

\end{document}